\documentstyle[epsfig]{aipproc}
\begin{document}
\title{Spin transfer in \\ high energy fragmentation processes}
\author{\underline{Liang Zuo-tang}, and Liu Chun-xiu}
\address{Department of Physics,
Shandong University,Jinan, Shandong 250100,China}
\maketitle

\begin{abstract}
We point out that measuring 
longitudinal polarizations of different 
hyperons produced in lepton induced reactions 
are ideal to study the spin transfer 
of the fragtmenting quark to produced hadron 
in high energy hadronization processes.
We briefly summarize 
the method used in calculating 
the hyperon polarizations in these processes,   
then present some of the results 
for $e^+e^-$ and $e^-p$ or $\nu p$ reactions
obtained using two different pictures for 
the spin structure of hyperon:
that drawn from polarized deep inelastic 
lepton-nucleon scattering data or 
that using SU(6) symmetric wave functions.
The results show in particular 
that measurements of such polarizations 
should provide useful information to the question 
of which picture is more suitable in describing 
the spin effects in the fragmentation processes. 
\end{abstract}

The talk was a summary of a series of papers [1-3] 
by C. Boros and ourselves. 
Spin transfer in high energy fragmentation process 
is defined as the probability for 
the polarization of the fragmenting quark to be transferred 
to the produced hadron. 
It is one of the important issue in connection with 
the spin effects in high energy fragmentation processes 
which have attracted much attention recently\cite{att}.
The problem contains the following two questions: 
(1) Will the polarization of the fragmenting quark 
be retained in the fragmentation process? 
(2) What is the relationship between 
the spin of the quark and that of the hadron 
which contains this quark?
Clearly, the answers to these questions 
depend not only on the hadronization mechanism 
but also on the spin structure of hadrons.
Study of such effects provide useful information 
for the spin structure of hadron and spin dependence 
of high energy reactions. 
There exist now two distinctively different 
pictures for the spin contents of the baryons:
the static quark model picture 
using SU(6) symmetric wave function  
[hereafter referred as SU(6) picture],  
and the picture drawn from the data for polarized deep inelastic 
lepton-nucleon scattering (DIS)
and SU(3) flavor symmetry in hyperon decay 
[hereafter referred as DIS picture].  
It is particular interesting to ask which picture 
is suitable to describe the relationship 
between the polarization of the fragmenting quark 
and that of the produced hadron which contains this quark. 
Obviously, the answer to this question 
is essential in the description of the 
puzzling hyperon transverse polarization 
observed already in the 1970s in unpolarized 
hadron-hadron reactions\cite{Hyp75}. 

It has been pointed out that\cite{BL98,LL20001} 
measurements of the longitudinal $\Lambda$ polarization 
in $e^+e^-$ annihilations at the $Z^0$ pole  
provide a very special check to the 
validity of SU(6) picture in connecting 
the spin of the constituent to the polarization of 
the hadron produced in the fragmentation processes. 
This is because the $\Lambda$ polarization in this case 
obtained from the SU(6) picture should be 
the maximum among different models. 
There are now data 
with reasonably high statistics available
from both ALEPH\cite{ALEPH96} and OPAL\cite{OPAL98} Collaborations. 
Their results show that the SU(6) picture seems 
to agree better with the data\cite{ALEPH96,OPAL98} 
compared with the DIS picture. 
This is rather surprising since the energy 
is very high at LEP thus the initial 
quarks and anti-quarks produced 
at the annihilation vertices of 
the initial $e^+e^-$ are certainly 
current quarks and current anti-quarks 
rather than the constituent quarks used 
in describing the static properties of hadrons 
using SU(6) symmetric wave functions.
It is thus interesting and instructive to  
make further checks in experiments by making 
complementary measurements.  
For this purpose, we have made a systematic study  
of hyperon polarizations in different lepton-induced 
reactions using the SU(6) or the DIS picture.  
The results we obtained can be used as further 
check of the pictures and now we give 
a brief summary of the 
calculation method and the obtained results.  

We first summarize the calculation method 
by taking $e^+e^-\to H_i+X$ as an example. 
  
Since the longitudinal polarization $P_{H_i}$ 
of the hyperon $H_i$ in the inclusive process  
$e^+e^-\to  H_i+X$ originates from 
the longitudinal polarization $P_f$  
of the initial quark $q^0_f$ 
(where the subscript $f$ denotes its flavor)
produced at the annihilation vertex 
of the initial state $e^+e^-$, 
we should consider the $H_i$'s 
which have the following different origins separately. 

(a) Hyperons which are directly produced 
and contain the 
initial quarks $q_f^0$'s originated from 
the annihilations of the initial $e^+$ and $e^-$;

(b) Hyperons which are decay products of other heavier 
hyperons which were polarized before their decay; 

(c) Hyperons which are directly produced but 
do not contain any initial quark $q_f^0$ from $e^+e^-$ 
annihilation;

(d) Hyperons which are decay products of other heavier hyperons 
which were unpolarized before their decay. 

It is clear that hyperons from (a) and (b) 
can be polarized while those from (c) and (d) are not. 
We obtain therefore, 
\begin{equation}
P_{H_i}={ {\sum\limits_f t^F_{H_i,f} P_f \langle n^a_{H_i,f}\rangle
+\sum\limits_{j} t^D_{H_i, H_j} P_{H_j} \langle n^b_{H_i, H_j}\rangle}
 \over
{\langle n^a_{H_i}\rangle +\langle n^b_{H_i}\rangle + 
\langle n^c_{H_i}\rangle +\langle n^d_{H_i}\rangle} }. 
\end{equation}
Here $P_f$ is the polarization of the initial quark $q_f^0$, 
and is determined by the electroweak vertex;
$\langle n^a_{H_i,f}\rangle$ is the average number of 
the hyperons which are directly produced and contain 
the initial quark of flavor $f$;
$\langle n^b_{H_i,H_j}\rangle$ is the average number of $H_i$ hyperons
coming from the decay of $H_j$ hyperons which are polarized;
$P_{H_j}$ is the polarization of the hyperon $H_j$   
before its decay;
$\langle n^a_{H_i}\rangle(\equiv \sum\limits_f \langle n^a_{H_i,f}\rangle$),
$\langle n^b_{H_i}\rangle(\equiv \sum\limits_j \langle n^b_{H_i,H_j}\rangle)$,
$\langle n^c_{H_i}\rangle$ and $\langle n^d_{H_i}\rangle$
are average numbers of hyperons in group (a), (b), (c) 
and (d) respectively; 
$t^F_{H_i,f}$ is the probability for 
the polarization of $q_f^0$ to be transferred 
to $H_i$ in the fragmentation process and 
is called the polarization transfer factor, 
where the superscript $F$ stands for fragmentation; 
$t^D_{H_i,H_j}$ is the probability for 
the polarization of $H_j$ to be transferred to 
$H_i$ in the decay process $H_j\to H_i+X$ and 
is called decay polarization transfer factor, 
where the superscript $D$ stands for decay. 
$t^F_{H_i,f}$ is equal to the fraction of 
spin carried by the $f$-flavor-quark 
divided by the average number of quark of flavor $f$ 
in the hyperon $H_i$.
This fractional contribution to the  
hyperon spin from $f$-flavor-quark 
is different in the above-mentioned 
SU(6) or the DIS picture. 
The results in the SU(6) picture can easily be obtained 
from the wave functions. 
In the DIS picture, the fractional contribution 
of quarks of different flavors 
to the spin of a baryon in the $J^P={1\over2}^+$ octet
is extracted from $\Gamma_1^p\equiv \int^1_0 g_1^p(x) dx$
obtained in deep-inelastic
lepton-proton scattering experiments
and the constants $F$ and $D$ obtained
from hyperon decay experiments. 
The way of doing this extraction 
is now in fact quite standard. 
A brief summay can, e.g., be found in the Appendix of [\ref{BL98}].
The results can be found e.g. in [\ref{BL98},\ref{LL20001}].
The decay polarization transfer factor $t^D_{H_i,H_j}$ 
is determined by the decay process and is independent 
of the process in which $H_j$ is produced.They can be 
extracted from the materials in Review of Particle 
Properties (see e.g. [\ref{GH93},\ref{BL98},\ref{LL20001}]). 

The average numbers of the hyperons of different 
origins mentioned above are determined 
by the hadronization mechanism and should be 
independent of the polarization of the initial quarks.
Hence, we can calculate them using a hadronization 
model which give a good description of the unpolarized 
data for multiparticle production in high energy reactions. 
Presently, such calculations can only be carried out using 
a Monte-Carlo event generator. 
We used the Lund string fragmentation model\cite{AGIS83}
implemented by JETSET in our calculations.

The method described above was first applied to 
$e^+e^-\to \Lambda +X$ at the $Z^0$ pole.
We recall that among all the $J^P={1\over 2}^+$ hyperons, 
$\Lambda$ is most copiously produced. 
Furthermore, 
the spin structure of $\Lambda$ in the $SU(6)$ picture
is very special, which makes it play a very special role 
in distinguishing the SU(6) and the DIS pictures. 
In the $SU(6)$ picture, 
spin of $\Lambda$ is completely carried by the $s$ valence quark,
while the $u$ and $d$ quarks have no contribution.
Since the initial $s$ quark produced 
in the annihilation of the initial $e^+e^-$ 
takes the maximum negative polarization,  
$|P_\Lambda|$ obtained using 
the SU(6) picture is the maximum among 
all the different models. 
In contrast, in the DIS picture, 
the $s$ quark carries only about $60\%$ of the $\Lambda$
spin, while the $u$ or $d$ quark each carries about $-20\%$.
The resulting $|P_\Lambda|$ should be substantially 
smaller than that obtained in the $SU(6)$ picture.
Comparing the maximum with experimental results 
provide us a good test of the validity of the picture. 

\begin{tabular}{lll}
\begin{minipage}[t]{6.6cm}
\psfig{file=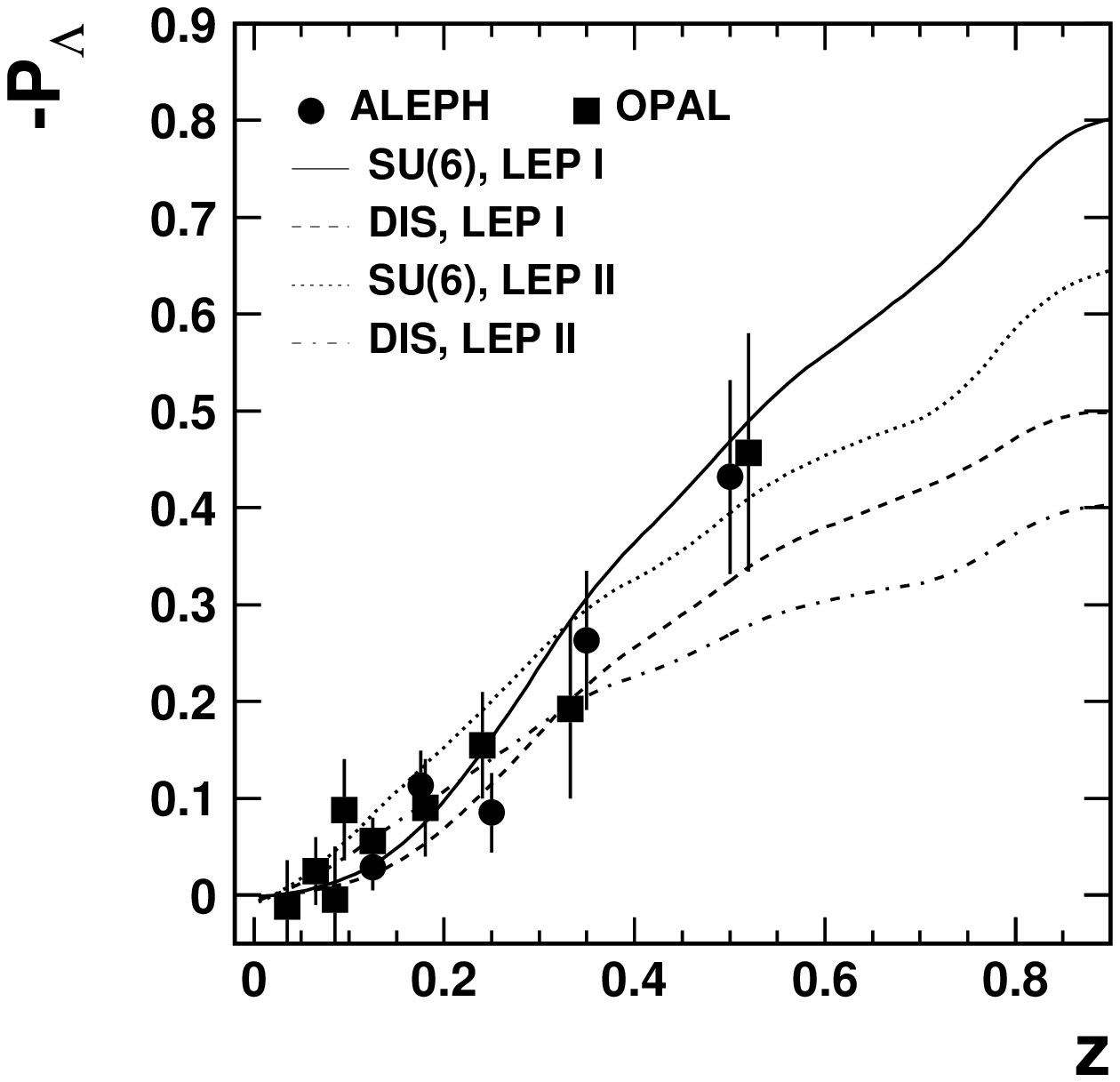,width=6cm}\\[-0.7cm]
{\small Fig.1: Longitudinal $\Lambda$ polarization, $P_\Lambda$,
in $e^+e^-\to \Lambda +X$ at LEP I and LEP II energies
as a function of $z$.
The data of ALEPH and those of OPAL are taken from
[\ref{ALEPH96}] and [\ref{OPAL98}] respectively.}\\
\end{minipage}
&
\begin{minipage}[t]{0.3cm}
\end{minipage}
&
\begin{minipage}[t]{6.6cm}
\psfig{file=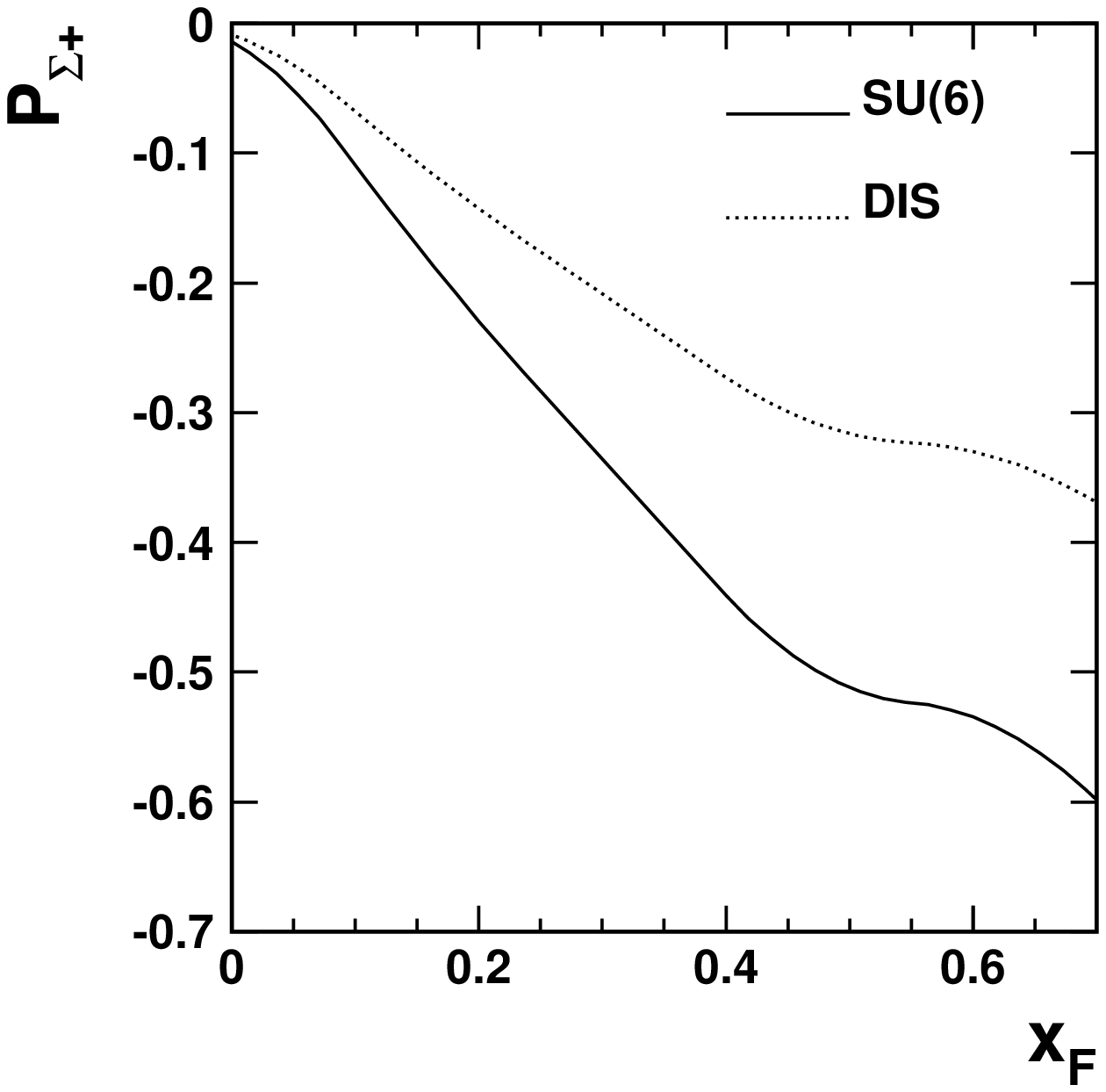,width=6cm}\\[-0.7cm]
{\small Fig.2: Longitudinal $\Sigma^+$ polarization, $P_{\Sigma^+}$,
in $\nu_\mu+p\to \mu^- +\Sigma^++X$ at $p_{inc}=500$GeV/c
as a function of $x_F$.}\\
\end{minipage} 
\\
\end{tabular}

Using the method described above, 
we obtained the longitudinal polarization of $\Lambda$ 
as shown in Fig.1.
A comparison with the ALEPH data \cite{ALEPH96} 
and the OPAL data \cite{OPAL98}
shows that the data\cite{ALEPH96,OPAL98} 
of both groups agree better with 
the calculated results based on the $SU(6)$ picture. 
But, these available data\cite{ALEPH96,OPAL98}  
are still far from accurate and enormous enough 
to make a decisive conclusion. 
Further complementary measurements are needed. 
We therefore made a systematic study of 
hyperon produced in different lepton-induced reactions 
and obtain in particular the following results 
which can be used as further checks of the pictures.

1.$\Lambda$ polarization in different subsamples of events 
in $e^+e^-\to \Lambda+X$.

We think it would be interesting to measure 
$\Lambda$ polarization in events where the following 
criteria are satisfied: 
(i) $\Lambda$ is the leading in one direction; 
(ii) the leading particle in the opposite direction is $K^+$. 
We expect that such $\Lambda$'s should mainly have the 
origin (a) mentioned above. 

Using the event generator JETSET, we showed that 
the $\Lambda$'s from (a) contribute 
indeed substantially higher in these events 
than they do in the average events 
and the obtained $|P_\Lambda|$ is also much higher [\ref{LL20001}]. 
This should be easily be check in experiments. 

2. Energy dependence of $P_\Lambda$ in $e^+e^-\to \Lambda X$.

To see the energy dependence, we calculated $P_\Lambda$ 
in $e^+e^-\to \Lambda X$ at LEP II energy. 
The results are also shown in Fig.1. 
We see a significant energy dependence due to that 
of $P_f$ of the initial quark $q_f^0$. 
This can also be checked.

3. Longitudinal polarization of other $J^P={1\over2}^+$ hyperons.

The production rates for other octet hyperons 
are smaller than that for $\Lambda$ so 
the statistic errors should be larger for 
the polarizations of these hyperons. 
On the other hand, decay contributions 
from heavier hyperons to these hyperons are also 
much less significant than that in case of $\Lambda$. 
Hence, the contaminations from the decay processes 
are much smaller. 
These conclusions can easily be checked using 
a Monte-Carlo event generator for $e^+e^-$ 
annihilation into hadrons.
On the other hand, 
we see also that the contribution 
from heavier hyperon decays is also much smaller.  
For example, for $\Sigma^+$'s, the decay contribution 
takes only about $7\%$ of the total rate. 
The situations for $\Sigma^-$, $\Xi^0$, and $\Xi^-$
are similar to that for $\Sigma^+$. 
Most of them are directly produced.
Hence, the theoretical uncertainties in the 
calculations for these hyperons are much smaller.
The study of polarizations of these hyperons 
should provide us with good complementary tests 
of different pictures.
We thus calculated 
the longitudinal polarizations of all these 
$J^P=(1/2)^+$ hyperons[\ref{LL20001}],
i.e., $\Sigma^+$, $\Sigma^-$, $\Xi^0$ and $\Xi^-$,
in $e^+e^-$ annihilation at LEP I and LEP II energies.
We found that they are all polarized and 
the polarizations are different for different hyperons. 
They are also different in SU(6) picture and the DIS picture 
which can indeed be used as complementary check to the picture.

4. Hyperon polarization in the current fragmentation 
region in lepton-nucleon deeply inelastic scatterings. 

The advantages of using hyperons in 
lepton-nucleon deeply inelastic scatterings are the following: 
First, we can study here not only longitudinal polarization transfer 
but also to check whether it is the same for transverse polarization case.
Second, flavor separation is, in some cases, automatically.
But it will be more difficult to reach the same statistics as 
that in $e^+e^-$ annihilation at the $Z^0$ pole.

We calculated the hyperon polarization in different 
reactions and different kinematic regions [3]. 
We found it is partcularly interesting that 
$\Lambda$ polarization in these reactions 
are usually small and also the differences from 
different pictures. 
On the other hand, polarizations of $\Sigma$ 
are larger and the differences between different pictures 
are also larger. 
Hence, it should be more sensitive to use $\Sigma$ 
as a check to different pictures than to use $\Lambda$ 
in these reactions. 
As an example, we show $P_\Sigma$ in 
$\nu_\mu+p\to\mu^-+\Sigma^++X$ in Fig.2.

This work was supported in part by the National Science Foundation of 
China and the Chinese Education Ministry.


\begin {thebibliography}{99}
\bibitem{att} See, e.g., the references we cited in 
        our publications [1-3].
\end{thebibliography}
\end{document}